# The role of traps in the photocurrent generation mechanism in thin InSe photodetectors

*Qinghua Zhao, Wei Wang, Felix Carrascoso-Plana, Wanqi Jie, Tao Wang\*, Andres Castellanos-Gomez\*, Riccardo Frisenda\**

Q. Zhao, Wei Wang, Prof. W. Jie, Prof. T. Wang
State Key Laboratory of Solidification Processing, Northwestern Polytechnical University, Xi'an, 710072, P. R. China
Key Laboratory of Radiation Detection Materials and Devices, Ministry of Industry and Information Technology, Xi'an, 710072, P. R. China
E-mail: taowang@nwpu.edu.cn

Q. Zhao, F. Carrascoso-Plana, Dr. R. Frisenda, Dr. A. Castellanos-Gomez
Materials Science Factory. Instituto de Ciencia de Materiales de Madrid (ICMM-CSIC), Madrid, E-28049, Spain.

E-mail: andres.castellanos@csic.es; riccardo.frisenda@csic.es

**Abstract:** Due to the excellent electrical transport properties and optoelectronic performance, thin indium selenide (InSe) has recently attracted attention in the field of 2D semiconducting materials. However, the mechanism behind the photocurrent generation in thin InSe photodetectors remains elusive. Here, we present a set of experiments aimed at explaining the strong scattering in the photoresponsivity values reported in the literature for thin InSe photodetectors. By performing optoelectronic measurements on thin InSe-based photodetectors operated under different environmental conditions we find that the photoresponsivity, the response time and the photocurrent power dependency are strongly correlated in this material. This observation indicates that the photogating effect plays an imporant role for thin InSe flakes, and it is the dominant mechanism in the ultra-high photoresponsivity of pristine InSe devices. In addition, when exposing the pristine InSe photodetectors to the ambient environment we observe a fast and irreversible change in the photoresponse, with a decrease in the photoresponsivity accompanied by an increase of the operating speed. We attribute this photodetector performance change (upon atmospheric exposure) to the decrease in the density of the traps present in InSe, due to the passivation of selenium vacancies by atmospheric oxygen species. This passivation is accompanied by a downward shift of the InSe Fermi level and by a decrease of the Fermi level pinning, which leads to an increase of the Schottky barrier between Au and InSe. Our study reveals the important role of traps induced by defects in tailoring the properties of devices based on 2D materials and offers a controllable route to design and functionalize thin InSe photodetectors to realize devices with either ultrahigh photoresposivity or fast operation speed.

**Introduction**

The isolation of ultrathin two-dimensional (2D) semiconducting materials, such as single-layer transition metal dichalcogenides (TMDCs) and few-layer black phosphorous (bP), has attracted large attention due to their potential applications in next-generation electronic and optoelectronic devices.[1-5] The large surface-to-volume ratio of these 2D materials, which on one side makes them very tunable and sensitive to external stimuli, on the other side can make these materials and the devices based on them extremely vulnerable to environmental degradation.[6-12] For example, when few-layer bP is exposed to the air, a fast degradation of the material occurs through a photooxidation process that leads to a reduction of the performances and the eventual failure of devices based on bP.[6-8] Similar environmental degradation phenomenon also has been observed on thin gallium selenide (GaSe),[9] gallium telluride (GaTe),[13] and even CVD (chemical vapor deposition)-grown single-layer $MoS_2$ and $WS_2$, two members of the TMDC family.[14] Moreover, the mobility of 2D semiconducting transistors kept in vacuum or encapsulated with boron nitride is typically more than one order of magnitude larger than the mobility measured in the air.[15-16] In the case of single-layer and bilayer $MoS_2$, among the various reports, the device mobility at room temperature ranges from 0.1 $cm^2V^{-1}s^{-1}$ in air to tens of $cm^2V^{-1}s^{-1}$ in vacuum or with top/bottom deposited protection materials.[17-19] All these observations can be explained by the presence of defects in the materials, such as chalcogen vacancies in the layered metal chalcogenides.[20-22] These defects can act as preferential sites for physical/chemical adsorption of environmental species (that can initiate the degradation process of 2D materials) and/or may introduce additional scattering of the carriers (that could act as harmful active traps in working devices).

Indium selenide (InSe), an n-type semiconductor which belongs to the III-VIA family, has recently attracted large attention because of its extraordinary charge transport properties, superior mechanical flexibility and strong light-matter interaction.[23-41] Various groups reported on transistors based on thin InSe fabricated on different substrates ($SiO_2$/Si, hexagonal Boron Nitride (h-BN), Poly(methyl methacrylate) (PMMA)) with mobility values as large as 3700 $cm^2V^{-1}s^{-1}$ at room temperature and ~13000 $cm^2V^{-1}s^{-1}$ at 4 K.[23, 25-28] The bandgap of 1.3 eV in bulk that becomes larger than 3 eV for an InSe single-layer makes this material interesting for broadband photodetection from the near-infrared to the near ultraviolet region of the electromagnetic spectrum.[30-40]

Various photodetectors based on thin InSe flakes (as the active channel part), including metal-semiconductor-metal (M-S-M) geometry and graphene based van der Waals heterostructures, have been reported in literature with responsivities going from 0.035 A $W^{-1}$ to ultrahigh values of ~$10^7$ A $W^{-1}$ and detectivities up to ~$10^{15}$ Jones [30-33, 35-38, 42] which are among the best performances reported for 2D photodetectors.[33, 40] See Supporting Information, Table S1, for a comparison between the reported values in the literature for the figures of merit of InSe photodetectors. Such a large scattering in the responsivity values has also been observed for other chalcogenide-based 2D materials. For example in $MoS_2$ photodetectors, traps states due to sulfur atomic vacancies influence the photocurrent generation giving rise to photogain and introducing Fermi level pinning at the metal-semiconductor (M-S) interface.[43-49] Similarly, in the case of InSe, the defects related to In adatoms and Se vacancies, which can be related to the In-rich atmosphere in which high-quality InSe crytals are grown,[50-51] may play important roles during the photocurrent generation process in thin InSe photodetectors.[20, 50-52] Interestingly, the presence of defects in InSe crystals, especially selenium vacancies ($V_{Se}$),[22] is predicted to promote the physical adsorption and chemical dissociation of $O_2$/$H_2O$ molecules and $V_{Se}$ can act as preferential sites for the adsorption of these molecules.[20, 52-55] This phenomenon can help in explaining the performance degradation reported in various works on thin InSe devices.[20, 33, 51, 54, 56] Recently, both theoretical calculations and experimental scanning transmission electron microscopy (STEM) reported that the selenium vacancies $V_{Se}$ in InSe crystals can be passivated by chemical dissociation of O atoms at these sites.[20, 22, 51] Similarly, Po-Hsun *et. al.* demonstrated that the change of Raman and X-ray photoelectron spectra (XPS) of pristine InSe flakes exposed to ambient air is comparable to that observed when exposing InSe to dry oxygen atmosphere. The authors attribute these observations to the formation of a surficial $InSe_{1-x}O_x$ layer that encapsulate the InSe beneath and promote a long-term stability of thin InSe devices.[26] Despite the large amount of work on InSe photodetectors, an important question that remains unanswered is how this passivation influences the optoelectronic properties of photodetectors based on thin InSe.

In this work, by comparing optoelectronic measurements obtained in InSe photodetectors operated under various environmental conditions, we propose a model to explain the photocurrent generation mechanism in InSe devices and how it is influenced by the environment. According to our observations, the optoelectronic properties of thin InSe photodetectors are stable over a long time of more than a few weeks when the devices are stored in vacuum condition. When exposing the devices to ambient conditions, we observe a fast decrease of the responsivity accompanied by an increase in the operation speed (reduction of the response time) of the device. The high photoresponsivity observed in pristine thin InSe photodetector can be attributed to a strong photogating effect, which mainly originates from traps for the minority carriers (holes). After exposing the device to the air, the passivation of the chalcogen vacancies $V_{Se}$ by oxygen induces a decrease in photocurrent, which can be explained by a decrease in the number of traps and a quenching of photogain and photogating in the system. Decreasing the density of traps in the 2D material has also a secondary effect on the devices, by reducing the Fermi level pinning we observe an increase of the Schottky barrier at the metal-semiconductor (M-S) interface consistent with the Schottky-Mott rule, as confirmed by scanning photocurrent measurements. Our results provide a further understanding of photocurrent generation mechanism in photodetectors based on thin InSe and can pave the way in utilizing this novel material in high-performance electronics and optoelectronics.

**Results and discussions**

Thin InSe photodetectors are fabricated by deterministic placement of InSe flakes (thicknesses going from ~5 nm to ~20 nm) mechanically exfoliated from an InSe bulk crystal grown by Bridgman method.[57] By characterizing the InSe crystal with transmission electronic microscope (TEM) and Raman spectroscopy, we find that the phase of the crystal is $\varepsilon$-type, see Figure S1 of the Supporting Information and our previous work.[36] Figure 1a shows the crystal structure of $\varepsilon$-InSe, with depicted two layers bound together by van der Waals forces. To isolate thin flakes of InSe, we first cleave larger flakes of the bulk InSe crystal onto Nitto tape (SPV 224) and then we transfer some of the flakes onto the surface of a polydimethylsiloxane (PDMS) stamp (Gel-Film by Gel-Pak) by gently pressing the tape against the stamp and peeling off slowly. The flakes on the PDMS can be then quickly transferred onto another arbitrary substrate with micrometric spatial precision by using a deterministic transfer setup.[58-59] Figure 1b shows a microscopic picture of a thin InSe flake isolated onto PDMS (left) and a picture of the final photodetecting device (right), formed by the same flake after being transferred bridging two pre-patterned electrodes (5 nm thick chromium sticking layer and 50 nm thick gold) evaporated on a $SiO_2$/Si substrate (with a $SiO_2$ thickness of 280 nm). To carry out the optoelectronic measurements the two gold electrodes (used a source and drain) are connected to a sourcemeter (Keithley® 2450) and the silicon substrate (heavily p-doped) is connected to a voltage source and used as a back gate. The differently colored regions of InSe deposited onto PDMS visible in Figure 1b are due to different thicknesses of the material. From atomic force microscopy (AFM) topography measurements, we find that the thinnest region of the flake, which forms the channel in the final device, is 9.1 nm thick corresponding to approximately 11 layers.[23]

Just after the fabrication of the device, we performed Raman spectroscopy measurements of the InSe flake. Figure 1c shows the Raman spectrum acquired on the region of the flake located above the gold electrode (indicated in Figure 1b). The spectrum shows three prominent peaks centered at 116 cm$^{-1}$, 178 cm$^{-1}$, and 227 cm$^{-1}$. These three peaks are due to vibrational modes of InSe and they can be attributed respectively to A'$_1$, E'' and A'$_1$. A fourth peak is visible at 200 cm$^{-1}$ and is due to A''$_2$ (see Figure S1b), which is sensitive to the crystalline phase.[36] Since the degradation of 2D materials is a common phenomenon, we test the stability of the InSe flake by repeating the Raman measurements after two weeks of exposure to ambient conditions. As can be seen from Figure 1c and Figure S2 the Raman spectra of the pristine flake and the one of the aged flake are very similar, the only difference being a ~15% reduction of the intensity of the peaks in the aged spectrum. This observation is consistent with previous reports from literature (see Supporting Information of Ref. 39)[39, 60] and we estimate that corresponds to the degradation of the top ~1-2 layers. Therefore, these Raman measurements indicate that the structure of mechanically exfoliated thin InSe flakes does not degrade completely upon exposure to ambient air, a different behavior from thin black phosporous or GaSe.[6, 9]

To investigate the properties of pristine InSe photodetectors, we perform optoelectronic measurements on a freshly prepared device kept in vacuum (sample #1 in Figure S3). We carried out the measurements in a home-made vacuum probe-station connected to a turbomolecular pumping station capable of reaching a base vacuum level of ~1 × 10$^{-6}$ mbar. All the measurements presented in this article were performed at room temperature ($T$ = 300 K). Figure 2a shows the transfer curve of the device, that is the source-drain current recorded while slowly changing the gate voltage $V_G$. This measurement was performed with a bias voltage $V_{DS}$ of 1 V and keeping the device in dark conditions. The device shows negligible current at negative gate voltages (off state, $V_G < V_T$) and starts to conduct current for positive voltages (on state, $V_G > V_T$). These observations indicate that the InSe channel is n-doped and from the plot we extract a threshold voltage $V_T$ = 10 V, which is in agreement with previous reports on InSe transistors.[23, 25-28] From this transfer curve we can also estimate the electron mobility using the equation:

$$\mu_n = \frac{L}{WC_{OX}V_{DS}}\left(\frac{dI_{DS}}{dV_G}\right), \tag{1}$$

where $L$ = 10 μm and $W$ = 24 μm are the channel length and width of the InSe photodetector, $C_{OX}$ = 115 μF m$^{-2}$ is the capacitance per unit area of 280 nm thick SiO$_2$ and $dI_{DS}/dV_G$ is the transconductance. Using the transconductance extracted from the linear regime of the transfer curve, we find a calculated mobility of ~0.06 cm$^2$V$^{-1}$s$^{-1}$ at a bias voltage $V_{DS}$ of 1 V. Compared with the reported results based on the InSe transistors fabricated by lithography method, this smaller value may be caused by a larger contact resistance in our case that does not involve the thermal evaporation of the electrodes onto the InSe flake.[59] In Figure 2b, the black curve shows the same transfer curve of panel a represented with a semi-logarithmic scale. From this graph, one can see that when passing from the off state to the on state the current flowing through the device increases of more than two orders of magnitude. The steepness of the current in the on-off transition region can be quantified by the subthreshold swing $S$, which is defined according to:

$$S = \frac{dV_G}{d(\log_{10}(I_{DS}/V_{DS}))}. \qquad (2)$$

From the data of Figure 2b, we find $S \sim 10$ V/decade, a value that is much larger than the ideal MOSFET (metal-oxide-semiconductor FET) subthreshold swing of 60 mV/decade and that is comparable to values reported previously for atomically thin high-gain photodetectors.[43] This large value for $S$ points to the existence of trap levels in the device.[61] Additional electrical measurements of this device can be found in Figure S4 of the Supporting Information.

After the characterization of the InSe photodetector in dark conditions, we test its optoelectronic properties and response to external illumination. The blue curve in Figure 2b corresponds to the transfer curve of the device under external illumination with a light source of wavelength 405 nm and power density of $10^{-4}$ W cm$^{-2}$. Under these conditions, the current passing through the device increases to a maximum of 1000 nA in the on state and to more than 200 nA in the off state, corresponding to an increase from the dark current of more than one order of magnitude in the on case and of three orders of magnitude in the off state. From the current measured while keeping the device in dark ($I_{dark}$) and under external illumination ($I_{light}$), we can calculate the responsivity $R$ and the detectivity $D^*$ of the photodetector using the following equations:[62-63]

$R = |I_{ph}| / (P\,A),$ \hfill (3)

and

$D^* = R\,A^{1/2} / (2\,e\,I_{dark})^{1/2},$ \hfill (4)

where $I_{ph}$ is the photocurrent calculated as ($I_{light} - I_{dark}$), $P$ is the external illumination density, $A$ is the active area of the photodetector (that we assume equal to the InSe channel area) and $e$ is the elementary charge. Note that the formula to calculate the detectivity of Eq. 4 assumes that the photodetector is limited by shot noise.[46] In Figure 2c we show these two figures of merit $R$ and $D^*$ as a function of gate voltage, calculated using the data of Figure 2b. The responsivity follows the same behavior of the photocurrent, reaching its maximum at positive gate voltages in the on state and decreasing when the transistor is in the off state. On the other hand, the detectivity is not monotonous and it reaches its maximum at a gate voltage of approximately -20 V and decreases for both large positive voltages and negative voltages. The values of *R* and *D\** reported here are comparable to the highest values reported in literature for thin InSe photodetectors, confirming the excellent performances of InSe as an active material in the fabrication of photodetectors.[33, 35] As a final comment, both the transfer curve of the device under illumination and the responsivity show a decrease when the device is in the off state. This can be explained by the presence of trap states for majority carriers near the conduction band of InSe whose occupation is controlled by the position of the Fermi level.[43] The measurements discussed up to this point, which have been performed in vacuum, indicate that trap levels are present in material and that these can influence the photodetection mechanism of InSe photodetector.

In the following, we focus on the influence of air on the performances of InSe photodetectors. Just after the device fabrication, a pristine InSe photodetector (sample #2 in Figure S3) was stored in high vacuum (pressure $\sim 10^{-6}$ mbar) and the optoelectronic performance was measured. To characterize the device in this case we studied its response in time to external square-wave modulated illumination. The red curve in Figure 3a shows the current flowing in the device at $V_{DS}$ = 1 V as a function of time (*I-t*) while switching on and off a 530 nm light source focused onto the surface of the device. The two curves of Figure 3a have been recorded with two illumination power densities, 450 mW µm$^{-2}$ and 225 mW µm$^{-2}$. When switching on the illumination at around 30 s with power 450 mW µm$^{-2}$ the current starts to slowly increase from ~400 pA to ~43 nA in approximately 100 s giving a responsivity of 0.09 A W$^{-1}$, while in the case of 225 mW µm$^{-2}$ the current reaches a value of ~25 nA and the responsivity is 0.11 A W$^{-1}$. This increase of the responsivity for decreasing illumination power densities points to the presence of traps for minority carriers (holes) in the device and we will discuss in more detail the power dependent measurements later in the article.[46-47]

The slow response of the photocurrent to external illumination shown in Figure 3a can be quantitatively characterized by the response time, which we estimate using the 10%-90% criterion. Here the response time is defined as the time that it takes for the current to increase from 10% of the saturation value to 90%. From the current vs. time traces (*I-t*) at 450 mW µm$^{-2}$ we estimate a rising time $\tau_{on}$ = 77 s and a decay time $\tau_{off}$ = 3.2 s. Figure 3b shows the data just discussed plotted in semi-logarithmic scale, where the small dark current of ~400 pA is visible. After carrying out the measurements in vacuum we exposed the device to air in ambient conditions for approximately 20 hours and then we repeated the *I-t* measurements. The green curve in Figure 3b is the corresponding *I-t* measured in air with the same illumination parameters used for the *I-t* in vacuum. As can be seen from the plot, both the dark current $I_{dark}$ and the current under illumination $I_{light}$ recorded in air conditions are much smaller than the initial values recorded in vacuum and the photocurrent $I_{ph}$ becomes ~1.2 nA, a ~40 times reduction. At the same time, we also observe a decrease of both the rising and decay time of the device, which becomes smaller than ~60 ms, approximately 2 or 3 orders of magnitude smaller than the initial values. Subsequently, after measuring the device in the air, we evacuated the chamber reaching again a pressure of ~$10^{-6}$ mbar and we carried out the same optoelectronic characterization. From the blue curve shown in Figure 3b, one can see that both $I_{dark}$ and $I_{light}$ remain at the same level recorded in the air indicating that the device underwent an irreversible transformation after being exposed to air for ~20 hours. The extracted responsivity is 0.002 A W$^{-1}$ and both the rising and the decay time are equal to ~40 ms. From these measurements, we find that by exposing the device to air, the InSe photodetector goes from an initial state in with larger responsivity but slower time response to a final state with smaller responsivity and faster time response.

In a photodetector, important information about the mechanism behind the photocurrent generation can be extracted from the power dependency of the photocurrent.[9, 64] During the *I-t* measurements discussed above, we also measured the photocurrent for different light intensities going from 45 mW µm$^{-2}$ to 450 mW µm$^{-2}$. For each value of light intensity, we extract the maximum photocurrent and we plot the results in Figure 3c. The photocurrent as a function of the illumination power density is shown in a log-log representation and the three datasets correspond to the three states of the device (**1** pristine in vacuum, **2** after 20 hours of exposure to air, **3** in vacuum after being exposed to air). From the graph, one can see that each of the three datasets follows approximately a straight line, indicating that the photocurrent and the illumination power are connected by a power law. This relation can be expressed by the equation:

$$I_{ph} = b \cdot P^{\alpha}, \tag{5}$$

where $\alpha$ is the dimensionless exponent of the power law and *b* is a parameter related to the photodetector responsivity. The value of the exponent $\alpha$ provides the information of traps present in the photodetecting system. In fact, in an ideal trap-free photodetector, the exponent $\alpha$ is equal to 1 meaning that the photocurrent scales linearly with the illumination power and the responsivity is constant as a function of power. When trap states (for minority carriers) are present in the system $\alpha$ becomes smaller than 1 and the responsivity depends sub-linearly on the illumination power (as for high powers most of the traps are already filled in and further illumination power cannot effectively increase the photogain), effectively decreasing for higher illumination powers. As can be seen from Figure 3c, by fitting the data to equation (5) we find that the InSe photodetector in its pristine state in vacuum (**1**) is characterized by $\alpha$ = 0.77. This value increases to 0.94 after that 20 hours of exposure of the device to air (**2**).

The final measurements performed on the device in vacuum after being exposed to air (**3**) show a value for $\alpha$ of 1.0. The observed evolution of $\alpha$ indicates that in pristine InSe photodetectors trap states play a role in the photocurrent generation process and that these trap states can be modified and eventually irreversibly removed after exposing the InSe photodetector to air.

The increase of $\alpha$ in InSe photodetectors exposed to air, discussed above for a single device, has been observed in all the five investigated Au-InSe-Au devices (see Figure S5 of the Supplementary Information). Moreover, a similar increase in $\alpha$ has been observed also in graphite-InSe-graphite devices (see supporting information Figure S7). Importantly, this common behavior is independent on the initial parameters of the measured devices such as responsivity, $\alpha$ or response time. This fact has been illustrated in Figure 4a-b where we collect the results of five different devices (shown in Figure S3) measured in the pristine state and after several exposure times to air. Figure 4a shows a semi-logarithmic graph of the decay time of the five devices plotted as a function of the exponent $\alpha$, extracted from power-dependent *I-t* measurements similar to the ones shown in Figure 3a. The different devices are represented by data-points with different shapes and colors. As can be seen from the statistical plot, among all the investigated InSe photodetectors the exponent $\alpha$ can take values between ~0.3 and 1 and correspondingly, the decay time goes from thousands of seconds when $\alpha$ is smaller than 0.5 (photogating dominated devices) to values smaller than 0.1 s (a value that could be limited by the experimental resolution of 0.04 s) when $\alpha$ tends to 1 (purely photoconductive devices). The plot shows a clear dependency of the decay time on the power exponent $\alpha$ for all the InSe photodetectors and indicates that the two variables are inversely proportional. A second statistical correlation can be observed between the responsivity $R$ and the decay time $\tau_{off}$ as shown in Figure 4b. In this log-log plot, the decay time of the five different InSe photodetectors is plotted versus the photoresponsivity extracted for each device. In this case, we measured responsivity values going from $10^{-3}$ A W$^{-1}$ to $10^{2}$ A W$^{-1}$ corresponding to decay times going from less than 0.1 s to more than 1000 s. The graph shows a correlation between the two with the data-points following approximately a straight line, which indicates that the two variables are connected by a power law. The black dashed line represents a power law with an exponent equal to 1, corresponding to a linear relationship between the responsivity and the response time. As can be seen from the plot, for large values of both $R$ and $\tau_{off}$, the data-points follow perfectly the linear relation, suggesting that the photogating effect is the dominant mechanism for InSe photodetectors with large responsivities. In fact, in photogating dominated devices the minority carriers get trapped in long-lived charge traps which limits in an effective way the response time of the device (they are typically slower than photoconductive devices) but provides an external source of photogain (when the channel drift time of the charge carriers is much shorter than the charge trapping time the device presents a photogain proportional to $\tau_{off}/\tau_{drift}$).[43, 46] For lower values of $R$ and $\tau_{off}$, the data-points start to deviate from the linear relation and present larger scattering, indicating that the photocurrent generation mechanism in these photodetectors is less dominated by photogating.

After exploring the correlation between $\alpha$, $R$ and $\tau_{off}$ we discuss a last set of experiments that highlight the change in the electronic structure of InSe after the exposure to air. After the initial optoelectronic characterization of device #1 in vacuum (transfer curves and photocurrent power dependency at 405 nm, see Figure S4 in the Supporting Information), we expose the device to air in dark environment for 2 hours and then we evacuate again the vacuum chamber and perform the optoelectronic measurements. We repeat these measurements during air-vacuum cycles until the performances of InSe photodetector become stable (approximately after 16 hours of air exposure). In Figure 4c we plot both the exponent $\alpha$ (left axis), extracted from the power dependency of the photocurrent, and the threshold voltage $V_T$ (right axis), extracted from the transfer curves, both plotted as a function of the exposure time to the air. As can be seen, the exponent $\alpha$ increases from 0.32 to 0.67, indicating the reduction in the trap states density and the change in the photocurrent generation mechanism from PG ($\alpha$ = 0.32) to PC ($\alpha$ = 0.67). At the same time $V_T$ (recorded in dark conditions) increases from ~10 V to ~50 V, a clear sign of the incremental p-doping of the InSe channel. The evolution of the transfer curves of this device and additional measurements under illumination can be found in Figure S6 of the Supplementary Information. In a second experiment, we study the evolution of the spatially resolved photocurrent in device #4 using scanning photocurrent microscopy. Briefly, in this technique we focus a 650 nm laser in a ~1 um² circular spot onto the surface of the InSe photodetector and move it across the sample while recording for each position the source-drain current and the intensity of the reflected light (see Figure S8 of the Supporting Information). Figure 4d shows the average current recorded in device #4 while scanning the laser spot from the source to the drain electrode in steps of 0.5 μm in pristine conditions (blue curves) and after 10 days of exposure to air (red). The line-profile of the pristine device at $V_{DS}$ = 1 V shows a broad and high photocurrent intensity region over the entire InSe channel area. On the contrary, the aged device presents a strong and sharp photocurrent peak located near the source contact at the end of the device channel, which is consistent with the presence of Schottky barriers at the InSe/gold contacts.[65-66] See figure S8 of the Supporting Information for the measurements under applied bias of $V_{SD}$ = -1 V.

Before discussing a model that can explain the photocurrent generation mechanism in thin InSe photodetectors it is instructive to summarize the experimental observations presented above. In summary, we studied five different InSe photodetectors in their freshly fabricated pristine state, and after the exposure to ambient conditions.

1. From electrical measurements in a field effect transistor configuration, we find that pristine InSe photodetectors are n-type doped (Figure 2a).
2. The optoelectronic figures-of-merit of pristine InSe photodetectors are excellent, with responsivities of the order of $10^3$ A W$^{-1}$ and detectivities of ~$10^{13}$ Jones. The responsivity depends on the gate voltage, decreasing when the InSe photodetectors are operated in the off state (Figure 2b-c).
3. From time-resolved measurements we find that the time response in pristine InSe photodetectors is slow and can be on the order of thousands of seconds (Figure 3a-b).
4. After exposing the InSe photodetectors to the air we observe a decrease of both the dark current and the current under illumination, accompanied by a decrease in response time (Figure 3b). Statistically, the responsivity and the response time across all the investigated photodetector are strongly correlated (Figure 4b).
5. The exposure to the air also affects the way in which the photocurrent depends on the illumination power density. While pristine InSe photodetectors show a sublinear dependency of $I_{ph}$ on $P$, with the power law exponent $α$ taking a value lower than 1, the exposure to air increase the linearity of this relation, with the power law exponent $α$ increasing toward 1 (Figure 3c).
6. Scanning photocurrent measurements (Figure 4d) shows that the interface between gold and pristine InSe shows lower Schottky barriers than that of aged InSe.

The n-type doping observed in the transfer curves of pristine thin InSe photodetectors measured in dark and in vacuum conditions has been observed in different experiments apart from the ones presented in this article.[23, 25-28] This behavior can be attributed to the presence of defects in the crystal such as selenium vacancies and indium interstial adatoms ($I_{In}$) that can be caused by the In-rich atmosphere in which high-quality InSe is grown.[50-51] These intrinsic defects are expected to modify the density of states of InSe and theoretical calculations predict the appearance of trap states linked to these defects. While $V_{Se}$ can introduce two sets of trap levels, the first one is located closer to the conduction band minimun (CBM) and the second is closer to the valence band maximum (VBM), the trap levels due to $I_{In}$ are located only closer to CBM.[51-52] In Figure 4e we show a schematic band diagram of the system composed by InSe, with the conduction band CB and the valence band VB displayed, and two gold electrodes acting as electrons reservoirs. The Fermi energy of the electrodes $E_F$ is closer to CB in agreement with the n-type conduction observed in the experiments. The majority carriers traps levels, which are closer to the conduction band, have a density $N_t$ and can act as pinning centers for the Fermi level of the device at zero bias. On the other hand, the minority carriers traps with density $P_t$ are expected to play a dominant role in the photocurrent generation process since they can generate photogain.[20, 51-52, 55] In dark conditions the InSe channel has a density of free electrons $n$ (located in CB) and of free holes $p$ (VB) and, as a consequence of the position of the Fermi level, there is a small density of trapped holes $p_t$ (meaning that most of the trap levels are occupied by electrons, density $n_t$). A free hole in VB (free electron in CB) has a certain probability of getting trapped by a trap level with rate $\tau_t^{-1}$. The opposite process also is possible and a trapped hole (electron) can jump from a trap level to VB (CB) with rate $\tau_d^{-1}$. In general, the trapping and detrapping rates $\tau_t^{-1}$ and $\tau_t^{-1}$ can be different and there can be also a difference between the rates relative to holes and electrons.

The band diagram of Figure 4e can be used to explain the current flowing through the device in dark conditions and serves as a basis to understand the photogeneration of current. When we illuminate the InSe photodetectors with external illumination, electron-hole pairs are generated and the density of free holes $p$ and free electrons $n$ increases. The separation of these electron-hole pairs due to the electric field related to the source-drain voltage gives rise to a photocurrent. This process that is usually called "photoconducting effect" (PC) and does not show photogain. In fact, the maximum responsivity achievable in a photodetector working solely by PC for illumination at 405 nm is ~0.33 A W$^{-1}$. In absence of active traps in the photodetector, PC is typically the dominant photocurrent generation mechanism. On the other hand, if hole traping levels are present in the system, each photogenerated hole has a certain probability (related to $\tau_t^{-1}$) of getting trapped in one of these levels for an average time $\tau_d$. This trapping process for minority carriers can give rise to "photogating effect" (PG), a process that can show photogain and can give responsivities much larger than ~0.33 A W$^{-1}$ at 405 nm. The magnitude of the photogain is related to the ratio between the trapping time and the drift time (the trapping time can be many orders of magnitude larger than the drift time).[43, 46] In the case of a pristine InSe photodetector measured in vacuum (stage **1**), the large responsivity values can be explained by PG. This mechanism also explains the large response times recorded for pristine photodetectors as in the PG mechanism the reponse speed is limitted by the trapping time. The power dependency of the photocurrent also confirms this scenario. In fact, a photodetector without trap levels is expected to be characterized by a value of the exponent $\alpha$ equal to 1 while in a photodetector containing active traps, $\alpha$ is smaller than 1. For the pristine InSe devices, dominant by PG effect, the interface between gold and InSe is characterized by small Schottky barriers, consistent with the scanning photocurrent measurements of device #4, and the alignment between the Fermi level of InSe and of the gold electrodes is determined by the Fermi level pinning to the trap levels.[49, 67]

The right panel of Figure 4e, schematise the evolution in the band structure after the exposure of InSe photodetectors to air (stage **2**). In this case, we propose that a change in the trap levels induce a strong change in the photocurrent generation process. In fact, the interaction between the selenium vacancies and the oxygen or water molecules present in air leads to a passivation of the defects, as predicted by theoretical calculations,[20, 51, 55] thus reducing the trap densities $N_t$ and $P_t$ (in our simplified scheme we removed completely the traps). The passivation process will eventually stop after all the avalaible vacancy sites are neutralized, characterized by a trap density much smaller than the initial density $P_t$ at stage **1**. At this stage of the evolution fewer holes can be trapped, effectively decreasing the photogain. The photodetector working principle shifts from PG to PC and the power dependency of the photocurrent illustrates well this evolution, with the aged InSe photodetectors showing a linear dependency of the photocurrent on the incident power. Once the InSe photodetectors reach this stage of evolution (stage **2**), they become much faster in responding to light (as the trapping time does not limit the photodetector speed anymore) but at the expense of a reduced responsivity. In this limit, the interface between gold and InSe is characterized by larger Schottky barriers and the alignment between InSe and gold can be calculated by the Schottky-Mott rule,[68-69] which predicts that the Schottky barrier formed at the interface has the value $\Phi_{SB,n} = W_M - X_S \sim 0.5$ eV, where $W_M$ is the work function of gold (~5.1 eV) and $X_S$ is the electron affinity of InSe (~4.6 eV). The Schottky barriers are visible in Figure 4d in the line-profile of the aged InSe photodetector.[65-66] Importantly, the final state of the photodetector seems to be stable both in the air and in vacuum suggesting that the passivation is a self-limiting process.[20] A final feature of the model, is the change of the relative alignement of the Fermi energy with respect to CB and VB during the evolution of InSe. In fact, the reduction of the density of trap levels is expected to induce a shift of the Fermi energy away from the conduction band, effectively introducing p-doping in the InSe channel.[52-53] The consequences of the p-doping can be observed in the reduction of the dark current in the air compared to the pristine one as shown in Figure 3b and in the shift of the treshold voltage.

**Conclusions**

In conclusion, by investigating the photocurrent generation process in InSe and its evolution in air/vacuum we find that traps levels can induce a strong photogain effect in pristine InSe photodetectors. The healing of these traps after exposure of InSe to air reduces the photogain and increases the operation speed of the photodetecting devices. The statistical analyses based on all our investigated thin InSe devices suggest that photoresponsivity, response time and the photocurrent scaling law exponent *α* are all strongly correlated quantities. The proposed band diagram model indicates that the high photoresponsivity in pristine thin InSe photodetectors can be at-tributed to the photogating effect, which mainly originates from the hole-trapping levels induced by selenium vacancies. The air-induced passivation of defects may reduce the traps density in thin InSe photodetectors, thus shifting the photocurrent generation mechanism from photogating to photoconductive. This change is accompanied by a decrease in photoresponse time and photoresponsivity as well as the growth of exponent $\alpha$. This work illustrates the strong effect of intrinsic defects, present in pristine 2D semiconductors, on their optoelectronic properties and how a controlled healing of these defects can be a powerful route to continuously tune the performance of these devices in a broad range of operational conditions.

**Materials and Methods**

**Sample fabrication.** Thin InSe flakes were fabricated with mechanical exfoliation method with scotch tape and Nitto tape (Nitto Denko® SPV 224) and then onto a polydimethylsiloxane substrate (PDMS). After inspection with optical microscopy transmission mode (Motic® BA310 MET-T), the selected flakes were transferred from the PDMS to the pre-patterned Au (50nm)/SiO$_2$ (280nm)/Si substrates (Osilla®) to fabricate the photodetector devices.[58-59] All the channel length of InSe photodetectors are 10 μm.

**Raman spectroscopy.** The Raman spectra of thin InSe flakes were obtained by a Bruker Senterra confocal Raman microscopy setup (Bruker Optik®, Ettlingen, Germany) with a laser excitation of 0.2 mW at 532 nm focused in a 1 μm diameter spot. The integration time is 20 s.

**Optoelectronic characterization.** Thin InSe photodetectors are characterized in a homebuilt high-vacuum (~1 × 10$^{-6}$ mbar, room temperature $T$ = 300 K) chamber. The electrical measurements (*I-V*, *I-t*) were performed with a source-meter source-measure unit (Keithley® 2450). The light source is provided by light emitting diodes (LEDD1B – T-Cube LED driver, Thorlabs®) with wavelength from 405 nm to 850 nm, coupled to a multimode optical fiber at the LED source and projected onto the sample surface by a zoom lens, creating a light spot on the sample with the diameter of 600 μm. The spatially resolved photocurrent maps under various source-drain voltage bias (-0.1 V, 0 V and 0.1 V) are acquired with an home-made scanning photocurrent system based on a modified microscope (BA310Met-H Trinocular, MOTIC) and a motorized XY scanning stage (8MTF, Standa).[70]

**AFM measurements.** The thickness of thin InSe flakes was measured by an ezAFM (by Nanomagnetics) atomic force microscope operated in dynamic mode. The cantilever used is Tap190Al-G by BudgetSensors with force constant 40 Nm$^{-1}$ and resonance frequency 300 kHz.

**ACKNOWLEDGEMENTS**

This is the authors' version (pre peer-review) of: Q Zhao, et al. Materials Horizons, 2020
https://doi.org/10.1039/C9MH01020C

This project has received funding from the European Research Council (ERC) under the European Union's Horizon 2020 research and innovation programme (grant agreement n° 755655, ERC-StG 2017 project 2D-TOPSENSE). EU Graphene Flagship funding (Grant Graphene Core 2, 785219) is acknowledged. RF acknowledges support from the Netherlands Organization for Scientific Research (NWO) through the research program Rubicon with project number 680-50-1515 and from the Spanish Ministry of Economy, Industry and Competitiveness through a Juan de la Cierva-formación fellowship (2017 FJCI-2017-32919). QHZ acknowledges the grant from China Scholarship Council (CSC) under No. 201700290035. TW acknowledges support from the National Natural Science Foundation of China: 51672216.

**COMPETING INTERESTS**

The authors declare no competing financial interests.

**FUNDING**

Netherlands Organization for Scientific Research (NWO): Rubicon 680-50-1515

Spanish Ministry of Economy, Industry and Competitiveness: Juan de la Cierva-formación fellowship 2017 FJCI-2017-32919

EU H2020 European Research Council (ERC): ERC-StG 2017 755655

EU Graphene Flagship: Grant Graphene Core 2, 785219

National Natural Science Foundation of China: 51672216

China Scholarship Council (CSC): 201700290035

**REFERENCES**

1. Wang, Q. H.; Kalantar-Zadeh, K.; Kis, A.; Coleman, J. N.; Strano, M. S., Electronics and optoelectronics of two-dimensional transition metal dichalcogenides. *Nature Nanotechnology* 2012**,** 7, 699-712.
2. Mak, K. F.; Shan, J., Photonics and optoelectronics of 2D semiconductor transition metal dichalcogenides. *Nature Photonics* 2016**,** 10, 216.
3. Li, L.; Yu, Y.; Ye, G. J.; Ge, Q.; Ou, X.; Wu, H.; Feng, D.; Chen, X. H.; Zhang, Y., Black phosphorus field-effect transistors. *Nature nanotechnology* 2014**,** 9, 372.
4. Castellanos-Gomez, A.; Vicarelli, L.; Prada, E.; Island, J. O.; Narasimha-Acharya, K.; Blanter, S. I.; Groenendijk, D. J.; Buscema, M.; Steele, G. A.; Alvarez, J., Isolation and characterization of few-layer black phosphorus. *2D Materials* 2014**,** 1, 025001.
5. Castellanos-Gomez, A., Why all the fuss about 2D semiconductors? *Nature Photonics* 2016**,** 10, 202-204.
6. Island, J. O.; Steele, G. A.; van der Zant, H. S.; Castellanos-Gomez, A., Environmental instability of few-layer black phosphorus. *2D Materials* 2015**,** 2, 011002.

7. Favron, A.; Gaufrès, E.; Fossard, F.; Phaneuf-L'Heureux, A.-L.; Tang, N. Y.; Lévesque, P. L.; Loiseau, A.; Leonelli, R.; Francoeur, S.; Martel, R., Photooxidation and quantum confinement effects in exfoliated black phosphorus. *Nature materials* 2015, 14, 826-832.
8. Wood, J. D.; Wells, S. A.; Jariwala, D.; Chen, K.-S.; Cho, E.; Sangwan, V. K.; Liu, X.; Lauhon, L. J.; Marks, T. J.; Hersam, M. C., Effective passivation of exfoliated black phosphorus transistors against ambient degradation. *Nano letters* 2014, 14, 6964-6970.
9. Zhao, Q.; Frisenda, R.; Gant, P.; Perez de Lara, D.; Munuera, C.; Garcia‐Hernandez, M.; Niu, Y.; Wang, T.; Jie, W.; Castellanos‐Gomez, A., Toward Air Stability of Thin GaSe Devices: Avoiding Environmental and Laser‐Induced Degradation by Encapsulation. *Advanced Functional Materials* 2018, 28, 1805304.
10. Wang, G.; Pandey, R.; Karna, S. P., Physics and chemistry of oxidation of two‐dimensional nanomaterials by molecular oxygen. *Wiley Interdisciplinary Reviews: Computational Molecular Science* 2017, 7.
11. Cao, W.; Jiang, J.; Xie, X.; Pal, A.; Chu, J. H.; Kang, J.; Banerjee, K., 2-D Layered Materials for Next-Generation Electronics: Opportunities and Challenges. *IEEE Transactions on Electron Devices* 2018, 65, 4109-4121.
12. Li, Q.; Zhou, Q.; Shi, L.; Chen, Q.; Wang, J., Recent advances in oxidation and degradation mechanisms of ultrathin 2D materials under ambient conditions and their passivation strategies. *Journal of Materials Chemistry A* 2019, 7, 4291-4312.
13. Yang, S.; Qin, Y.; Chen, B.; Özçelik, V. O.; White, C. E.; Shen, Y.; Yang, S.; Tongay, S., Novel surface molecular functionalization route to enhance environmental stability of tellurium-containing 2D layers. *ACS applied materials & interfaces* 2017, 9, 44625-44631.
14. Gao, J.; Li, B.; Tan, J.; Chow, P.; Lu, T.-M.; Koratkar, N., Aging of transition metal dichalcogenide monolayers. *ACS nano* 2016, 10, 2628-2635.
15. Qiu, H.; Xu, T.; Wang, Z.; Ren, W.; Nan, H.; Ni, Z.; Chen, Q.; Yuan, S.; Miao, F.; Song, F., Hopping transport through defect-induced localized states in molybdenum disulphide. *Nature communications* 2013, 4, 2642.
16. Park, W.; Park, J.; Jang, J.; Lee, H.; Jeong, H.; Cho, K.; Hong, S.; Lee, T., Oxygen environmental and passivation effects on molybdenum disulfide field effect transistors. *Nanotechnology* 2013, 24, 095202.
17. Radisavljevic, B.; Radenovic, A.; Brivio, J.; Giacometti, i. V.; Kis, A., Single-layer MoS 2 transistors. *Nature nanotechnology* 2011, 6, 147.
18. Fuhrer, M. S.; Hone, J., Measurement of mobility in dual-gated MoS 2 transistors. *Nature nanotechnology* 2013, 8, 146.
19. Chhowalla, M.; Jena, D.; Zhang, H., Two-dimensional semiconductors for transistors. *Nature Reviews Materials* 2016, 1, 16052.
20. Liu, Y.; Stradins, P.; Wei, S. H., Air Passivation of Chalcogen Vacancies in Two‐Dimensional Semiconductors. *Angewandte Chemie* 2016, 128, 977-980.
21. Zhou, W.; Zou, X.; Najmaei, S.; Liu, Z.; Shi, Y.; Kong, J.; Lou, J.; Ajayan, P. M.; Yakobson, B. I.; Idrobo, J.-C., Intrinsic structural defects in monolayer molybdenum disulfide. *Nano letters* 2013, 13, 2615-2622.
22. Hopkinson, D. G.; Zólyomi, V.; Rooney, A. P.; Clark, N.; Terry, D. J.; Hamer, M.; Lewis, D. J.; Allen, C. S.; Kirkland, A. I.; Andreev, Y., Formation and Healing of Defects in Atomically Thin GaSe and InSe. *ACS nano* 2019.
23. Bandurin, D. A.; Tyurnina, A. V.; Geliang, L. Y.; Mishchenko, A.; Zólyomi, V.; Morozov, S. V.; Kumar, R. K.; Gorbachev, R. V.; Kudrynskyi, Z. R.; Pezzini, S., High electron mobility, quantum Hall effect and anomalous optical response in atomically thin InSe. *Nature nanotechnology* 2017, 12, 223.
24. Mudd, G. W.; Svatek, S. A.; Ren, T.; Patanè, A.; Makarovsky, O.; Eaves, L.; Beton, P. H.; Kovalyuk, Z. D.; Lashkarev, G. V.; Kudrynskyi, Z. R., Tuning the bandgap of exfoliated InSe nanosheets by quantum confinement. *Advanced Materials* 2013, 25, 5714-5718.
25. Li, M.; Lin, C. Y.; Yang, S. H.; Chang, Y. M.; Chang, J. K.; Yang, F. S.; Zhong, C.; Jian, W. B.; Lien, C. H.; Ho, C. H., High Mobilities in Layered InSe Transistors with Indium‐Encapsulation‐Induced Surface Charge Doping. *Advanced Materials* 2018, 1803690.
26. Ho, P.-H.; Chang, Y.-R.; Chu, Y.-C.; Li, M.-K.; Tsai, C.-A.; Wang, W.-H.; Ho, C.-H.; Chen, C.-W.; Chiu, P.-W., High-mobility InSe transistors: the role of surface oxides. *ACS nano* 2017, 11, 7362-7370.
27. Sucharitakul, S.; Goble, N. J.; Kumar, U. R.; Sankar, R.; Bogorad, Z. A.; Chou, F.-C.; Chen, Y.-T.; Gao, X. P., Intrinsic electron mobility exceeding 103 cm2/(V s) in multilayer InSe FETs. *Nano letters* 2015, 15, 3815-3819.
28. Huang, Y.-T.; Chen, Y.-H.; Ho, Y.-J.; Huang, S.-W.; Chang, Y.-R.; Watanabe, K.; Taniguchi, T.; Chiu, H.-C.; Liang, C.-T.; Sankar, R., High-Performance InSe Transistors with Ohmic Contact Enabled by Nonrectifying Barrier-Type Indium Electrodes. *ACS applied materials & interfaces* 2018, 10, 33450-33456.
29. Liu, X.; Ren, J.-C.; Zhang, S.; Fuentes-Cabrera, M.; Li, S.; Liu, W., Ultrahigh Conductivity in Two-Dimensional InSe via Remote Doping at Room Temperature. *The journal of physical chemistry letters* 2018, 9, 3897-3903.
30. Lei, S.; Wen, F.; Ge, L.; Najmaei, S.; George, A.; Gong, Y.; Gao, W.; Jin, Z.; Li, B.; Lou, J., An atomically layered InSe avalanche photodetector. *Nano letters* 2015, 15, 3048-3055.
31. Lei, S.; Ge, L.; Najmaei, S.; George, A.; Kappera, R.; Lou, J.; Chhowalla, M.; Yamaguchi, H.; Gupta, G.; Vajtai, R., Evolution of the electronic band structure and efficient photo-detection in atomic layers of InSe. *ACS nano* 2014, 8, 1263-1272.
32. Tamalampudi, S. R.; Lu, Y.-Y.; Kumar U, R.; Sankar, R.; Liao, C.-D.; Moorthy B, K.; Cheng, C.-H.; Chou, F. C.; Chen, Y.-T., High performance and bendable few-layered InSe photodetectors with broad spectral response. *Nano letters* 2014, 14, 2800-2806.
33. Wells, S. A.; Henning, A.; Gish, J. T.; Sangwan, V. K.; Lauhon, L. J.; Hersam, M. C., Suppressing Ambient Degradation of Exfoliated InSe Nanosheet Devices via Seeded Atomic Layer Deposition Encapsulation. *Nano letters* 2018.

34. Chang, Y.-R.; Ho, P.-H.; Wen, C.-Y.; Chen, T.-P.; Li, S.-S.; Wang, J.-Y.; Li, M.-K.; Tsai, C.-A.; Sankar, R.; Wang, W.-H., Surface oxidation doping to enhance photogenerated carrier separation efficiency for ultrahigh gain indium selenide photodetector. *ACS Photonics* 2017, 4, 2930-2936.
35. Feng, W.; Wu, J.-B.; Li, X.; Zheng, W.; Zhou, X.; Xiao, K.; Cao, W.; Yang, B.; Idrobo, J.-C.; Basile, L., Ultrahigh photo-responsivity and detectivity in multilayer InSe nanosheets phototransistors with broadband response. *Journal of Materials Chemistry C* 2015, 3, 7022-7028.
36. Yang, Z.; Jie, W.; Mak, C.-H.; Lin, S.; Lin, H.; Yang, X.; Yan, F.; Lau, S. P.; Hao, J., Wafer-scale synthesis of high-quality semiconducting two-dimensional layered InSe with broadband photoresponse. *ACS nano* 2017, 11, 4225-4236.
37. Luo, W.; Cao, Y.; Hu, P.; Cai, K.; Feng, Q.; Yan, F.; Yan, T.; Zhang, X.; Wang, K., Gate Tuning of High‐Performance InSe‐Based Photodetectors Using Graphene Electrodes. *Advanced Optical Materials* 2015, 3, 1418-1423.
38. Mudd, G. W.; Svatek, S. A.; Hague, L.; Makarovsky, O.; Kudrynskyi, Z. R.; Mellor, C. J.; Beton, P. H.; Eaves, L.; Novoselov, K. S.; Kovalyuk, Z. D., High broad‐band photoresponsivity of mechanically formed InSe–graphene van der Waals heterostructures. *Advanced Materials* 2015, 27, 3760-3766.
39. Chen, Z.; Biscaras, J.; Shukla, A., A high performance graphene/few-layer InSe photo-detector. *Nanoscale* 2015, 7, 5981-5986.
40. Kang, J.; Wells, S. A.; Sangwan, V. K.; Lam, D.; Liu, X.; Luxa, J.; Sofer, Z.; Hersam, M. C., Solution‐Based Processing of Optoelectronically Active Indium Selenide. *Advanced Materials* 2018, 30, 1802990.
41. Zhao, Q.; Frisenda, R.; Wang, T.; Castellanos-Gomez, A., InSe: a two-dimensional semiconductor with superior flexibility. *Nanoscale* 2019, 11, 9845-9850.
42. Yang, H.-W.; Hsieh, H.-F.; Chen, R.-S.; Ho, C.-H.; Lee, K.-Y.; Chao, L.-C., Ultraefficient Ultraviolet and Visible Light Sensing and Ohmic Contacts in High-Mobility InSe Nanoflake Photodetectors Fabricated by the Focused Ion Beam Technique. *ACS applied materials & interfaces* 2018, 10, 5740-5749.
43. Furchi, M. M.; Polyushkin, D. K.; Pospischil, A.; Mueller, T., Mechanisms of photoconductivity in atomically thin MoS2. *Nano letters* 2014, 14, 6165-6170.
44. Tsai, D.-S.; Liu, K.-K.; Lien, D.-H.; Tsai, M.-L.; Kang, C.-F.; Lin, C.-A.; Li, L.-J.; He, J.-H., Few-layer MoS2 with high broadband photogain and fast optical switching for use in harsh environments. *Acs Nano* 2013, 7, 3905-3911.
45. Zhang, W.; Huang, J. K.; Chen, C. H.; Chang, Y. H.; Cheng, Y. J.; Li, L. J., High‐gain phototransistors based on a CVD MoS2 monolayer. *Advanced materials* 2013, 25, 3456-3461.
46. Buscema, M.; Island, J. O.; Groenendijk, D. J.; Blanter, S. I.; Steele, G. A.; van der Zant, H. S.; Castellanos-Gomez, A., Photocurrent generation with two-dimensional van der Waals semiconductors. *Chemical Society Reviews* 2015, 44, 3691-3718.
47. Huo, N.; Konstantatos, G., Recent Progress and Future Prospects of 2D‐Based Photodetectors. *Advanced Materials* 2018, 1801164.
48. Kim, C.; Moon, I.; Lee, D.; Choi, M. S.; Ahmed, F.; Nam, S.; Cho, Y.; Shin, H.-J.; Park, S.; Yoo, W. J., Fermi level pinning at electrical metal contacts of monolayer molybdenum dichalcogenides. *ACS nano* 2017, 11, 1588-1596.
49. Schulman, D. S.; Arnold, A. J.; Das, S., Contact engineering for 2D materials and devices. *Chemical Society Reviews* 2018, 47, 3037-3058.
50. Wang, H.; Shi, J.-j.; Huang, P.; Ding, Y.-m.; Wu, M.; Cen, Y.-l.; Yu, T., Origin of n-type conductivity in two-dimensional InSe: In atoms from surface adsorption and van der Waals gap. *Physica E: Low-dimensional Systems and Nanostructures* 2018, 98, 66-73.
51. Xiao, K.; Carvalho, A.; Neto, A. C., Defects and oxidation resilience in InSe. *Physical Review B* 2017, 96, 054112.
52. Kistanov, A. A.; Cai, Y.; Zhou, K.; Dmitriev, S. V.; Zhang, Y.-W., Atomic-scale mechanisms of defect-and light-induced oxidation and degradation of InSe. *Journal of Materials Chemistry C* 2018, 6, 518-525.
53. Politano, A.; Chiarello, G.; Samnakay, R.; Liu, G.; Gürbulak, B.; Duman, S.; Balandin, A.; Boukhvalov, D., The influence of chemical reactivity of surface defects on ambient-stable InSe-based nanodevices. *Nanoscale* 2016, 8, 8474-8479.
54. Shi, L.; Zhou, Q.; Zhao, Y.; Ouyang, Y.; Ling, C.; Li, Q.; Wang, J., Oxidation mechanism and protection strategy of ultrathin Indium Selenide: Insight from Theory. *The journal of physical chemistry letters* 2017, 8, 4368-4373.
55. Ma, D.; Li, T.; Yuan, D.; He, C.; Lu, Z.; Lu, Z.; Yang, Z.; Wang, Y., The role of the intrinsic Se and In vacancies in the interaction of O2 and H2O molecules with the InSe monolayer. *Applied Surface Science* 2018, 434, 215-227.
56. Wei, X.; Dong, C.; Xu, A.; Li, X.; Macdonald, D. D., Oxygen-induced degradation of the electronic properties of thin-layer InSe. *Physical Chemistry Chemical Physics* 2018, 20, 2238-2250.
57. Ishii, T., High quality single crystal growth of layered InSe semiconductor by Bridgman technique. *Journal of crystal growth* 1988, 89, 459-462.
58. Castellanos-Gomez, A.; Buscema, M.; Molenaar, R.; Singh, V.; Janssen, L.; van der Zant, H. S.; Steele, G. A., Deterministic transfer of two-dimensional materials by all-dry viscoelastic stamping. *2D Materials* 2014, 1, 011002.
59. Frisenda, R.; Navarro-Moratalla, E.; Gant, P.; De Lara, D. P.; Jarillo-Herrero, P.; Gorbachev, R. V.; Castellanos-Gomez, A., Recent progress in the assembly of nanodevices and van der Waals heterostructures by deterministic placement of 2D materials. *Chemical Society Reviews* 2018, 47, 53-68.
60. Pozo-Zamudio, D.; Schwarz, S.; Klein, J.; Schofield, R.; Chekhovich, E.; Ceylan, O.; Margapoti, E.; Dmitriev, A.; Lashkarev, G.; Borisenko, D., Photoluminescence and Raman investigation of stability of InSe and GaSe thin films. *arXiv preprint arXiv:1506.05619* 2015.

61. Ayari, A.; Cobas, E.; Ogundadegbe, O.; Fuhrer, M. S., Realization and electrical characterization of ultrathin crystals of layered transition-metal dichalcogenides. *Journal of applied physics* 2007, 101, 014507.
62. Choi, W.; Cho, M. Y.; Konar, A.; Lee, J. H.; Cha, G. B.; Hong, S. C.; Kim, S.; Kim, J.; Jena, D.; Joo, J., High‐detectivity multilayer MoS2 phototransistors with spectral response from ultraviolet to infrared. *Advanced materials* 2012, 24, 5832-5836.
63. Choi, M. S.; Qu, D.; Lee, D.; Liu, X.; Watanabe, K.; Taniguchi, T.; Yoo, W. J., Lateral MoS2 p–n junction formed by chemical doping for use in high-performance optoelectronics. *ACS nano* 2014, 8, 9332-9340.
64. Lopez-Sanchez, O.; Lembke, D.; Kayci, M.; Radenovic, A.; Kis, A., Ultrasensitive photodetectors based on monolayer MoS 2. *Nature nanotechnology* 2013, 8, 497.
65. Jin, H.; Li, J.; Wan, L.; Dai, Y.; Wei, Y.; Guo, H., Ohmic contact in monolayer InSe-metal interface. *2D Materials* 2017, 4, 025116.
66. Shi, B.; Wang, Y.; Li, J.; Zhang, X.; Yan, J.; Liu, S.; Yang, J.; Pan, Y.; Zhang, H.; Yang, J., n-Type Ohmic contact and p-type Schottky contact of monolayer InSe transistors. *Physical Chemistry Chemical Physics* 2018, 20, 24641-24651.
67. Ho, C. H.; Chu, Y. J., Bending photoluminescence and surface photovoltaic effect on multilayer InSe 2D microplate crystals. *Advanced Optical Materials* 2015, 3, 1750-1758.
68. Liu, Y.; Guo, J.; Zhu, E.; Liao, L.; Lee, S.-J.; Ding, M.; Shakir, I.; Gambin, V.; Huang, Y.; Duan, X., Approaching the Schottky–Mott limit in van der Waals metal–semiconductor junctions. *Nature* 2018, 557, 696.
69. Sze, S. M.; Ng, K. K., *Physics of semiconductor devices*. John wiley & sons: 2006.
70. Reuter, C.; Frisenda, R.; Lin, D. Y.; Ko, T. S.; Perez de Lara, D.; Castellanos‐Gomez, A., A versatile scanning photocurrent mapping system to characterize optoelectronic devices based on 2D materials. *Small Methods* 2017, 1, 1700119.

**FIGURES**

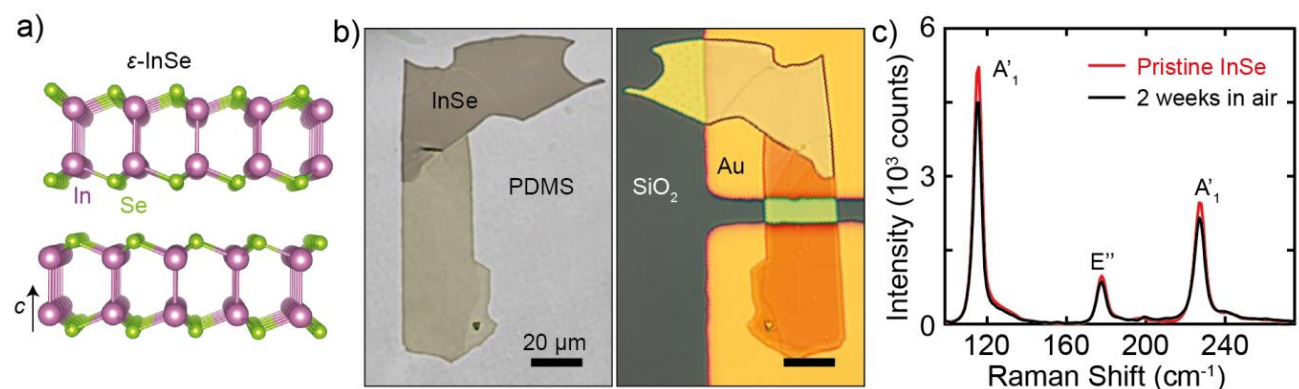

**Figure 1**: **Layered semiconducting ε-InSe.** a) Crystal structure of InSe. b) Optical pictures of few-layer InSe exfoliated onto PDMS (left) and deterministically transferred onto pre-patterned gold electrodes (right). c) Raman spectra (532 nm laser, power density 0.11 mWµm$^{-2}$) of the InSe flake shown in panel b recorded in the pristine state and after two weeks in air conditions.

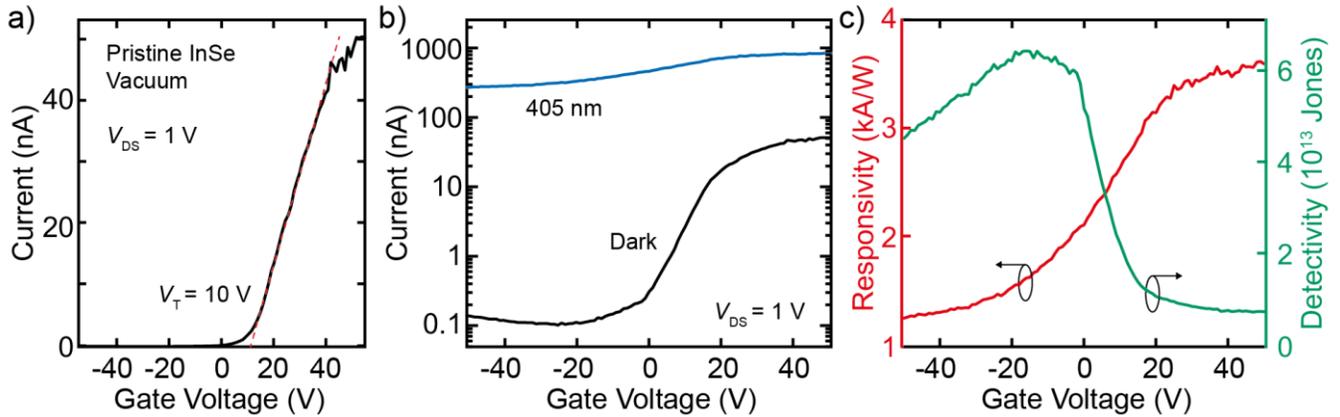

**Figure 2: Optoelectronic characterizations of a pristine InSe photodetector (# 1) in vacuum.** a) Current versus back gate voltage transfer curve at $V_{SD}$ = 1 V of a pristine InSe photodetector kept in vacuum and in dark. b) Current versus gate voltage in dark and under illumination at 405 nm (power density 0.92 W m$^{-2}$) represented in semi-logarithmic scale. c) Responsivity ($R$, left axis) and detectivity ($D^*$, right axis) of pristine InSe extracted from the curves in panel (b).

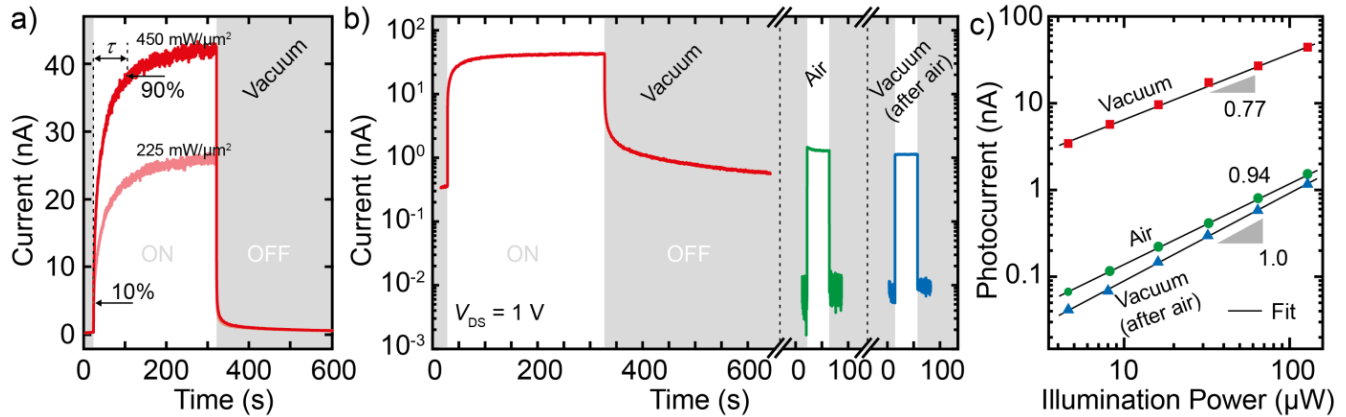

**Figure 3**: **Evolution in the air of an InSe photodetector (# 2).** a) Current at $V_{SD}$ = 1 V recorded in an InSe photodetector as a function of time while turning on and off a 530 nm light source with two different illumination powers. b) Current at $V_{SD}$ = 1 V recorded in an InSe photodetector as a function of time while turning on and off illumination at 530 nm (power density 450 mW/um$^{-2}$). The device was kept first in vacuum (pressure ~10$^{-6}$ mbar), then it was transferred into the air for ~20 hours (~10$^3$ mbar) and finally it was kept again in vacuum. c) Photocurrent as a function of illumination power at 530 nm recorded in the three environments of the panel (b).

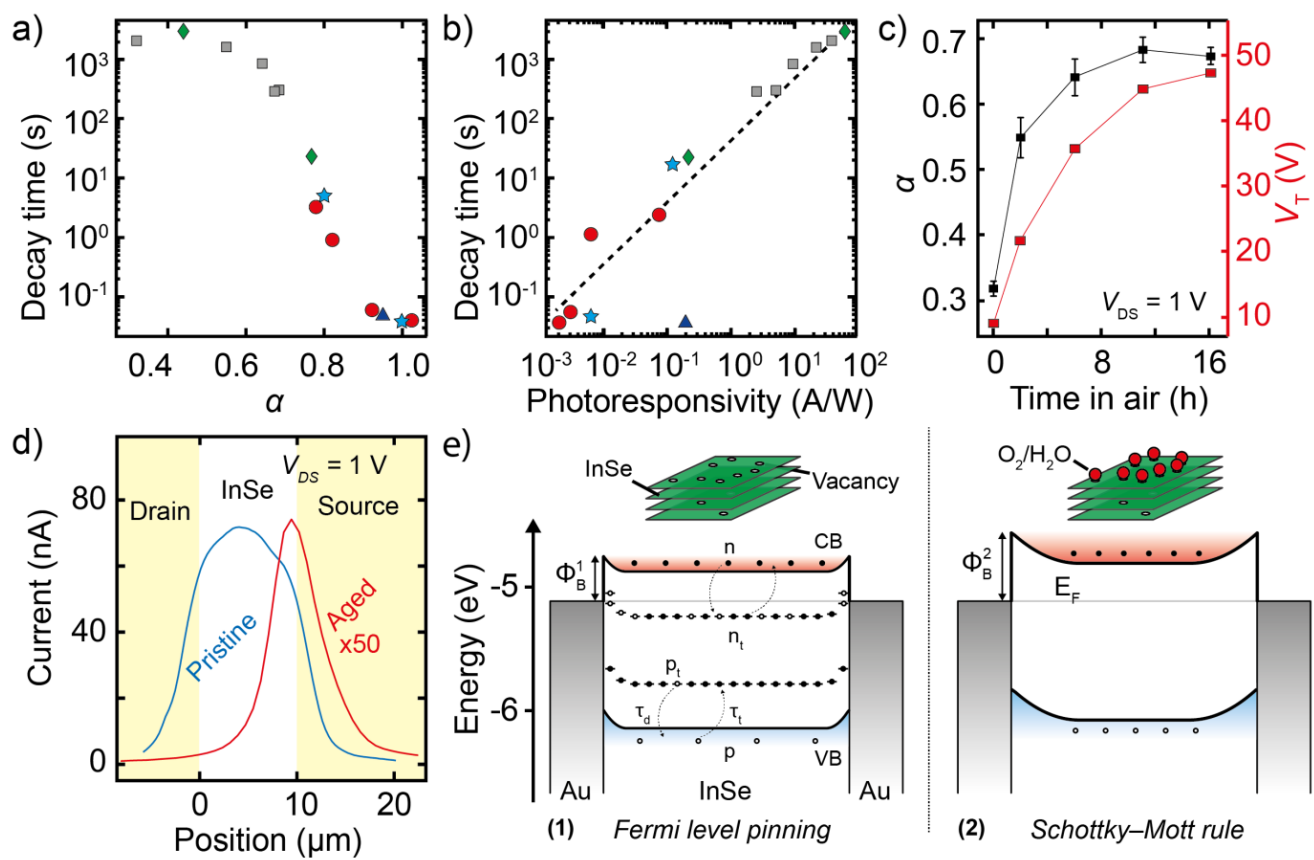

**Figure 4**: **Photocurrent generation mechanism in the InSe photodetectors.** a) Decay time as a function of the exponent $\alpha$ of the photocurrent-power scaling law extracted from all the investigated InSe photoetectors. b) The relationship between decay time and photoresponsivity in various InSe phtdetectors. All the dots in the same shape and color were measured with the same device after different air exposure times. c) Photocurrent exponent $\alpha$ (left axis) and threshold voltage $V_T$ (right axis) recorded in a device (#1) exposed to air as a function of time. d) Spatially resolved photocurrent at $V_{ds}$ = 1 V of device #4 recorded just after fabrication (blue) and after 10 days in air (red). e) Schematic band diagram of InSe in dark conditions with traps and Fermi level pinning (**1**, left) and in absence of traps with the alignment predicted by the Schottky-Mott rule (**2**, right). Depicted there are the valence (VB) and conduction band (CB) of InSe, the gold elecrodes and the Fermi energy ($E_F$) and a set of hole trapping levels (whose density is $P_t$) and electron trapping level (density $N_t$). The density of free electrons, free holes, trapped electrons and trapped holes are respectively n, p, $n_t$. and $p_t$.

## Supplementary Information:

## The role of traps in the photocurrent generation mechanism in thin InSe photodetectors

*Qinghua Zhao, Wei Wang, Felix Carrascoso-Plana, Wanqi Jie, Tao Wang\*, Andres Castellanos-Gomez\*, Riccardo Frisenda\**

Q. Zhao, Wei Wang, Prof. W. Jie, Prof. T. Wang
State Key Laboratory of Solidification Processing, Northwestern Polytechnical University, Xi'an, 710072, P. R. China
Key Laboratory of Radiation Detection Materials and Devices, Ministry of Industry and Information Technology, Xi'an, 710072, P. R. China
E-mail: taowang@nwpu.edu.cn

Q. Zhao, F. Carrascoso-Plana, Dr. R. Frisenda, Dr. A. Castellanos-Gomez
Materials Science Factory. Instituto de Ciencia de Materiales de Madrid (ICMM-CSIC), Madrid, E-28049, Spain.

E-mail: andres.castellanos@csic.es; riccardo.frisenda@csic.es

**Table S1: The photoresponsivity and response time of various thin InSe photodetectors with metal-InSe-metal geometry found in literature.**

| Reference | Device geometry | Wavelength | Response time | Responsivity |
|---|---|---|---|---|
| 30 | **Al-InSe-Al** | **543 nm** | **87 µs** | **---** |
| 31 | **Metal-InSe-metal** | **532 nm** | **488 µs** | **34.7 mA W$^{-1}$** |
| 32 | **Cr/Au-InSe-Cr/Au** | **450 nm** | **50 ms** | **6.9 – 157 A W$^{-1}$** |
| 33 | **Cr/Au-InSe-Cr/Au** | **515.6 nm** | **---** | **~ 10$^7$ A W$^{-1}$** |
| 35 | **Cr/Au-InSe-Cr/Au** | **700 nm** | **5 ms** | **~ 10$^4$ A W$^{-1}$** |
| 36 | **Au-InSe-Au** | **370 nm** | **0.5 s** | **27 A W$^{-1}$** |
| 37 | **Ti/Au-InSe-Ti/Au** | **500 nm** | **5.63 s** | **700 A W$^{-1}$** |
| 42 | **Pt/Au/Pt** | **325 nm** | **---** | **~ 10$^7$ A W$^{-1}$** |
| 37 | **G-InSe-G** | **500 nm** | **120 µs** | **60 A W$^{-1}$** |
| 38 | **G-InSe-G** | **633 nm** | **1 ms** | **4000 A W$^{-1}$** |

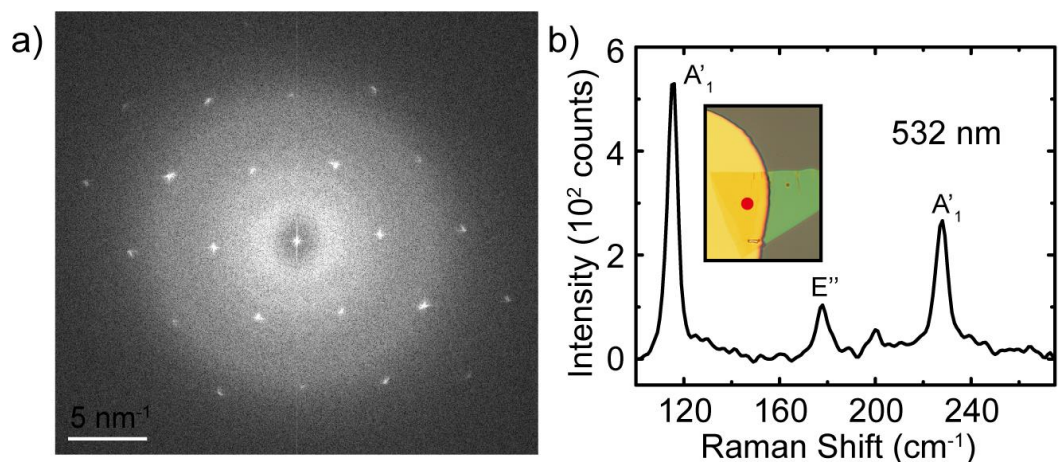

**Figure S1:** Crystal structure characterizations of thin InSe flakes. (a) The diffraction pattern of thin InSe flake by transmission electron microscope (TEM). The hexagonal geometry of the pattern indicates that the InSe is β or ε

phase. (b) The Raman spectra of thin InSe flake on Au. The peak centered at 200 cm$^{-1}$ gives the confirmation that the InSe is ε phase.

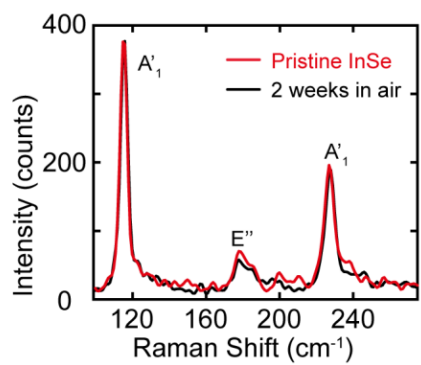

**Figure S2:** Raman spectra (532 nm laser, power density 0.11 mW µm$^{-2}$) of the InSe flake shown in Figure 1(b) on SiO$_2$/Si substrate recorded in the pristine state and after two weeks in air conditions. The Raman characterization on SiO$_2$/Si and Au/SiO$_2$/Si indicates that the thin InSe flake in the air is structurally stable.

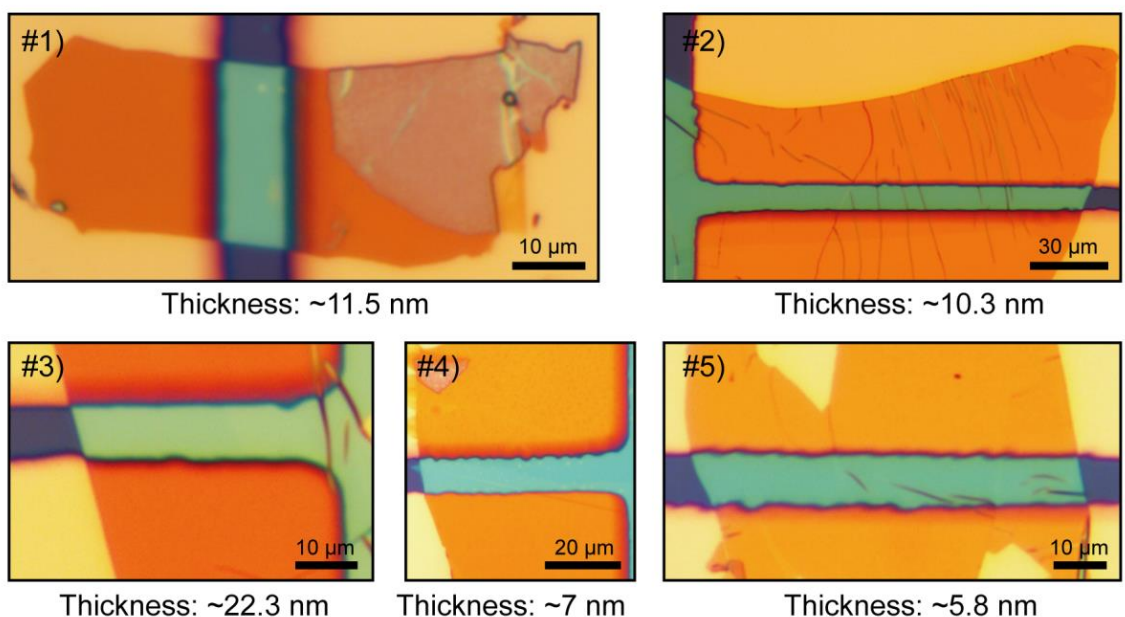

**Figure S3:** Optical pictures of five thin InSe photodetectors with various thicknesses and labeled #1 to #5 investigated in this work.

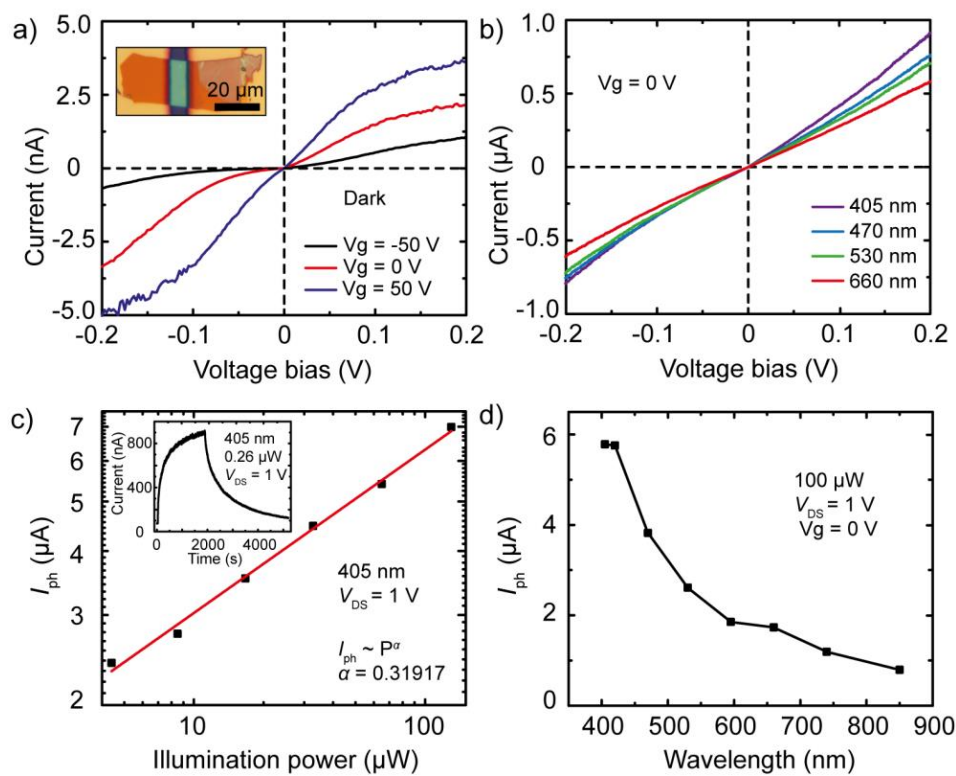

**Figure S4:** The optoelectronic characterization in vacuum (0.5-×10$^{-6}$ mbar - 1-×10$^{-6}$ mbar) of the InSe photodetector (#1). (a) The *I-V* curves of the InSe photodetector in dark condition with the gate voltage $V_g$ = -50 V, 0 V, and 50 V, respectively. (b) The *I-V* curves of the InSe photodetector under various illumination wavelength with the same power density of 354 W m$^{-2}$. (c) The photocurrent versus illumination intensity (~ 4 – 128 μW) plotted in log-log scale with the 405 nm illumination. The inset figure is an *I*-t curve for the InSe photodetector. (d) The photocurrent evolution with different wavelength under the illumination density of 354 W m$^{-2}$ at $V_{DS}$ = 1 V.

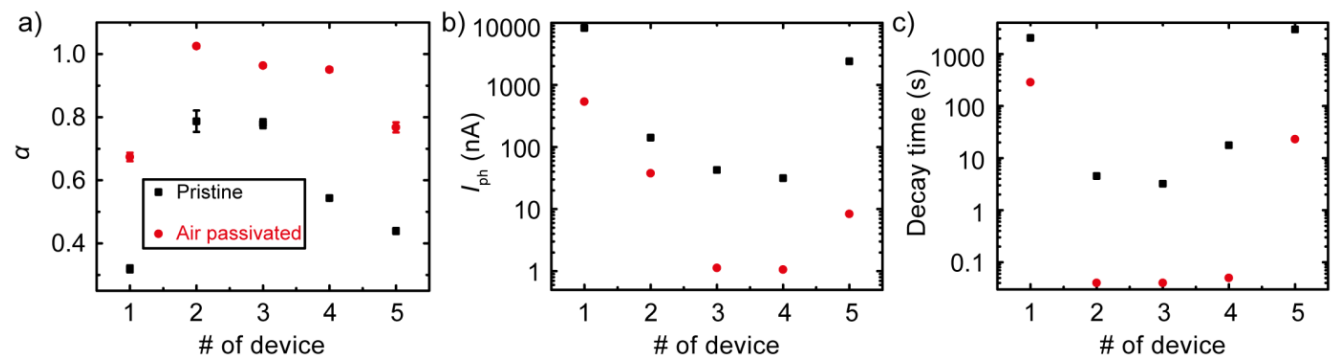

**Figure S5:** Statistic results of five InSe photodetectors (in Figure S3) before and after air passivation. $\alpha$ (a), photocurrent (b), and decay time (c) of five InSe photodetectors with pristine state and after air passivation under the same measurement conditions. All the tested InSe photodetectors share the same manner when exposed to air.

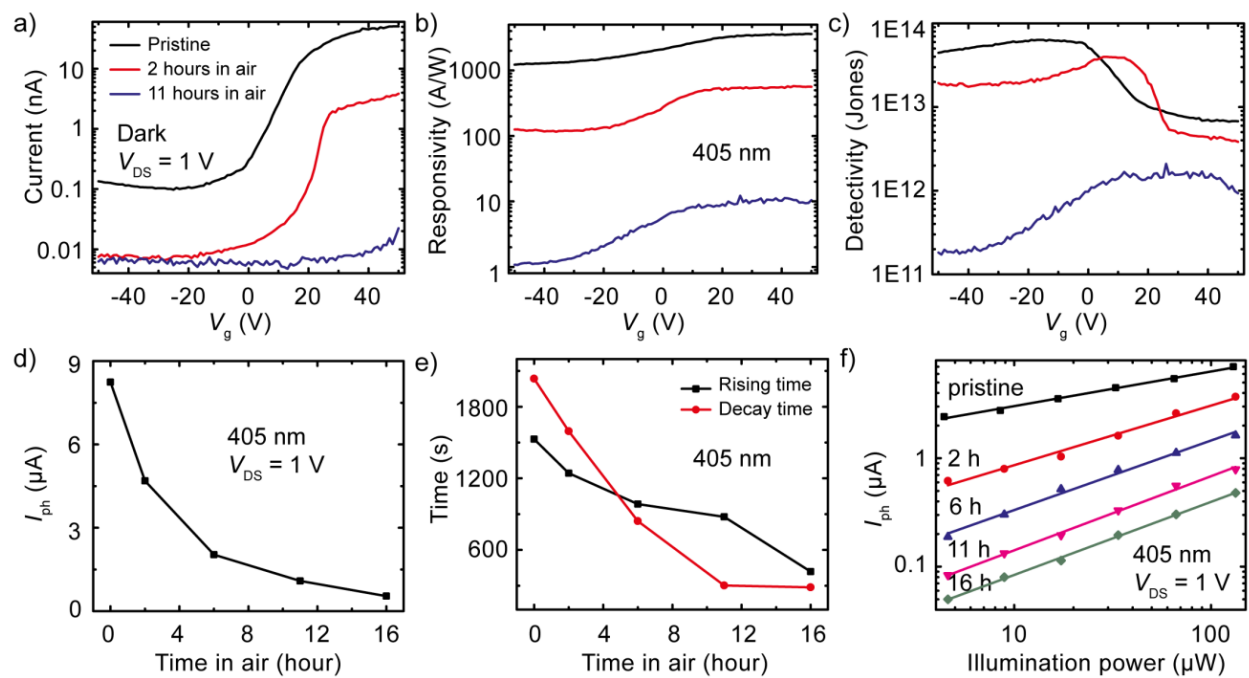

**Figure S6:** Time-dependent performance of the InSe photodetector (#1) when exposed to air. The gate-dependence of the InSe photodetector in dark (a), photoresponsivity (b) and detectivity (c) of 405 nm at 0.92 W m$^{-2}$ illumination after exposed to air 0 h, 2 h, 11 h. The evolution of photocurrent (at 906 W m$^{-2}$), rising and decay time (at 906 W m$^{-2}$), and photocurrent – illumination intensity dependence of 405 nm light as a function of exposure time in air.

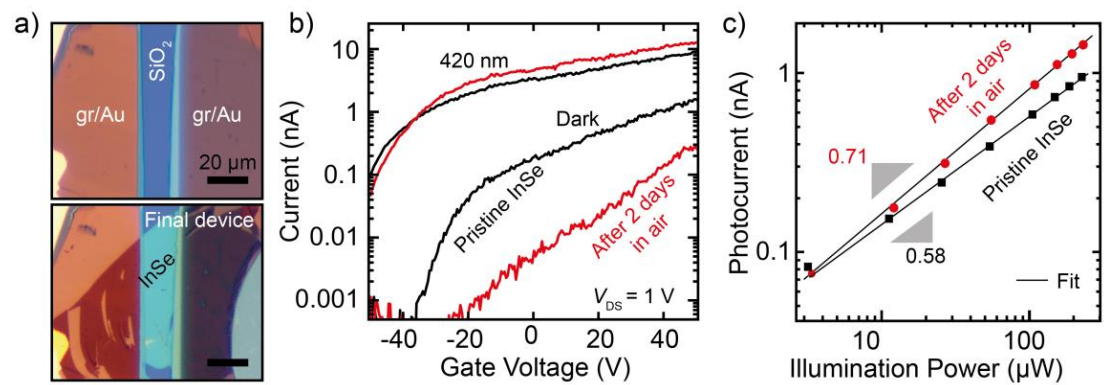

**Figure S7:** Time-dependent performance of a graphite-InSe-graphite photodetector when exposed to air.

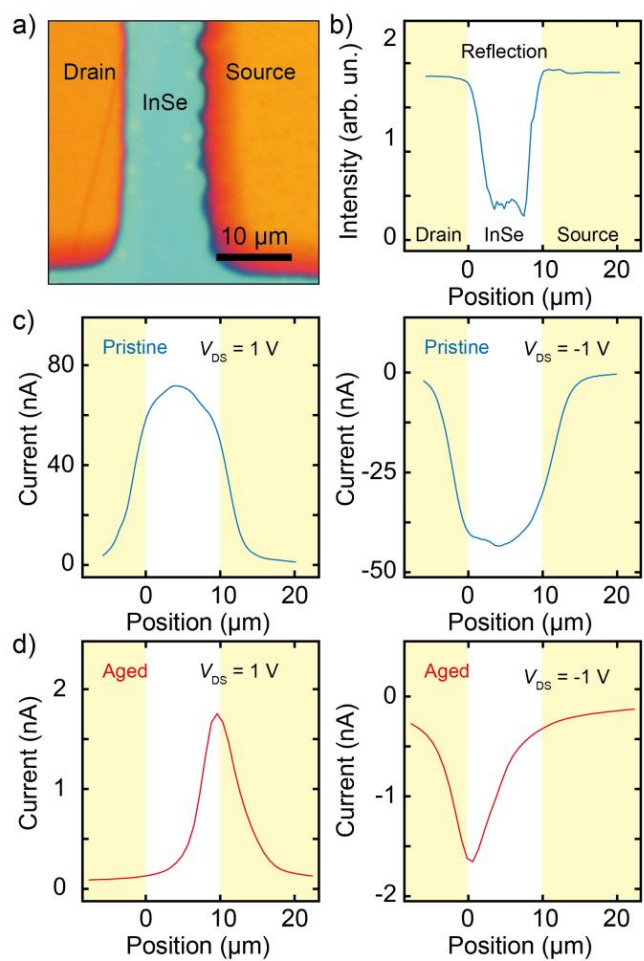

**Figure S8:** a) Optical image of device #4. b) Intensity of the laser reflection from the sample recorded during the scanning photocurrent measurements of panels c and d. c-d) Spatially resolved photocurrent at $V_{ds}$ = ±1 V of device #4 recorded just after fabrication (blue) and after 10 days in air (red).

**TOC graphics:**

# The role of traps in the photocurrent generation mechanism in thin InSe photodetectors

*Qinghua Zhao, Wei Wang, Felix Carrascoso-Plana, Wanqi Jie, Tao Wang\*, Andres Castellanos-Gomez\*, Riccardo Frisenda\**

Q. Zhao, Wei Wang, Prof. W. Jie, Prof. T. Wang
State Key Laboratory of Solidification Processing, Northwestern Polytechnical University, Xi'an, 710072, P. R. China
Key Laboratory of Radiation Detection Materials and Devices, Ministry of Industry and Information Technology, Xi'an, 710072, P. R. China
E-mail: taowang@nwpu.edu.cn

Q. Zhao, F. Carrascoso-Plana, Dr. R. Frisenda, Dr. A. Castellanos-Gomez
Materials Science Factory. Instituto de Ciencia de Materiales de Madrid (ICMM-CSIC), Madrid, E-28049, Spain.

E-mail: andres.castellanos@csic.es; riccardo.frisenda@csic.es

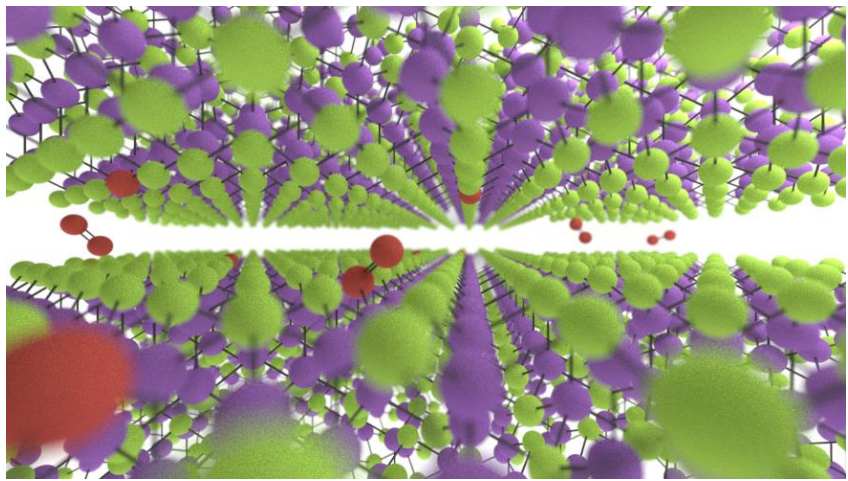